# Speaker Identification by GMM based *i* Vector

Soumen Kanrar
Department of Computer Science
Vidyasagar University
Midnapour, West Bengal, India
rscs_soumen@mail.vidyasagar.ac.in

## ABSTRACT
Speaker Identification process is to identify a particular vocal cord from a set of existing speakers. In the speaker identification processes, unknown speaker voice sample targets each of the existing speakers present in the system and gives a predication. The predication may be more than one existing known speaker voice and is very close to the unknown speaker voice. The model is a Gaussian mixture model built by the extracted acoustic feature vectors from voice. The i-vector based dimension compression mapping function of the channel depended speaker, and super vector give better predicted scores according to cosine distance scoring associated with the order pair of speakers. In the order pair, the first coordinate is the unknown speaker i.e. test speaker, and the second coordinates is the existing known speaker i.e. target speaker. This paper presents the enhancement of the prediction based on i- vector in compare to the normalized set of predicted score. In the simulation, known speaker voices are collected through different channels and in different languages. In the testing, the GMM voice models, and GMM based i-Vector speaker voice models of the known speakers are used among the numbers of clusters in the test data set.

## Keywords
Speaker Identification, Gaussian mixture mode(GMM), Acoustic feature vectors, Decision threshold, i-Vector.

## 1. INTRODUCTION
The acoustic signal corresponding to articulation is independent of language. Text independent and language independent voices are identified by the tracking of the vocal tract. The human utterances described in terms of a sequence of segments, and on the further crucial assumption that each segment can be characterized by an articulatory target [10]. 'Articulation' is the activity of the vocal organs in making a speech sounds. The aforesaid biometrics offers greater potentiality over the traditional methods in person recognition by GMM based speaker identification [1,2,3,4].In particular, voice recognition technology produces relatively low to medium error rate, and it has a high public acceptance rate due to not noticeable nature of voice sample. In general, the voice is the acoustic signal characteristic of a person's individual articulation. The articulation is the aspect of pronunciation involving the articulatory organs. So the identification of the unknown speaker is to identify the vocal track of the speaker from the number of existing speaker model present in the system [10]. Vocal cords produce acoustic energy by vibrating as air passes between then. If the claim speaker voice is very near to an existing model in the system or numbers of models, the initial stage of testing basically drills with the one to many matching. The numerical score gives the best probable prediction about the speaker with the list of the voice model present in the voice recognition system. We observe a large amount overlap between the impostors and the true speaker model in the speaker recognition system. It increases the equal error rate higher. The score normalization reduces the equal error rate by the z-norm and T-norm [8].The decision threshold is very crucial to maintain the false accept and false reject [9].The decision threshold truly depends on the environment from where the voice is collected i.e. the collected voice sample is fully contaminated or less contaminated or pure and clean with white- noise [7]. On the other hand, GMM based i-Vector gives cosine based score predication according to Bhattacharyya postulates. The paper organized as section one gives brief description of the problem. Section two presents the architecture of the identification procedure. The model description is presented in the section three and four. Section five and six are described about the simulation and conclusion remarks.

## 2. SYSTEM ARCHITECTURE
The initial procedure for Gaussian mixture model is presented in the flow diagram 1.The Gaussian mixture model is created through a number of steps. The input acoustic signal has been considered with the sampling rate of 4000, 8000 and 16000. Sampling rate is a number of sample per unit time. 'Quantization and Sampling' convert the analog signal to digital. Sampling rate is the number of samples per unit of time taken from the continuous signal to make a discrete signal.

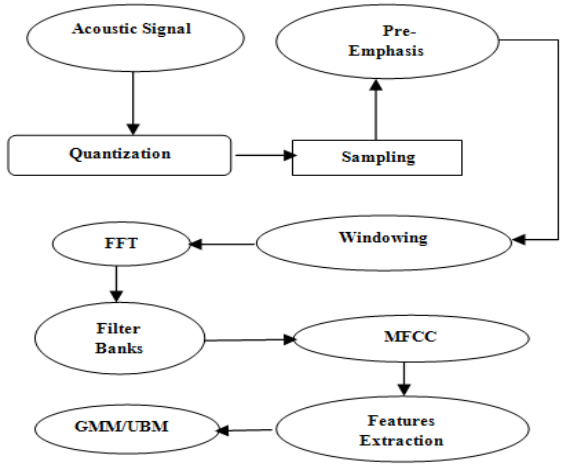

**Figure 1. GMM Creation Flow Diagram**

Pre Emphasis filters that higher frequency, in Quantization compact the domain. Windowing extracts the spectral feature by running the frame. The Hamming window is used, which shrinks the value of the signal towards zero at the window boundaries, to avoiding discontinuities. FFT (Fast Fourier Transformation) a way to analyzing the spectral properties of the input signal in the frequency domain. Filter Bank is the class of methods that process on multiple frequency bands of a given signal. It is a series of band pass frequency filters which are multiplied one by one with the spectrum in order to get an average value in a particular frequency band. The acoustic feature is extracted from the MFCC (Mel Frequency Cepstral Coefficient) [6] according to the figure 1. Melody's frequency Cepstral coefficients (MFCC) are collectively built up by the individual Melody frequency Cepstral (MFC). MFC is a physical representation of the short term power spectrum of an acoustic signal in a particular frequency band on a linear cosine transform of the log power spectrum [11]. The extracted acoustic feature from the voice signal after normalization produces various acoustic classes. These acoustic classes belong to an individual speaker voice or a set of speakers. The GMM is the soft representation of the various acoustic classes of an individual person's voice or a set of speakers. The probability of a feature vector of being in the acoustic classes is represented by the mixture of different Gaussian probability distribution functions. The GMM training model or the universal background model (UBM) is performed by the maximum a posterior adaption of a set of means $M_i$. The size of the set is nX1024, n=1,2,3.. . Now, nX1024 is the collected number of distinct voice sample from the population of well balanced, male and female to create the UBM.

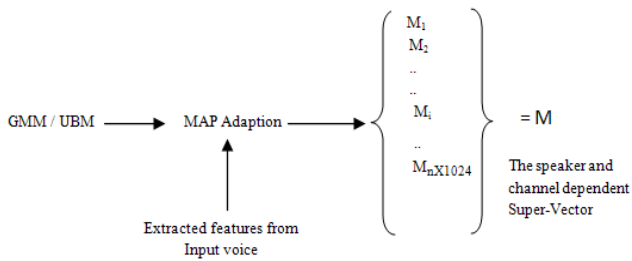

**Figure 2. Speakers and Channel Dependent Super Vector**

The super vector $'M'$ according to figure '2' is representing mapping between utterance and the high dimension vector space. So $'M'$ is a speaker and channel dependent super vector of concatenated GMM. The Joint Factor analysis [16]–[17] a speaker utterance is represented by a super-vector that consists of additive components from a speaker and a channel. Due to the linearity of Gaussian mixture model (GMM) [12] and generative equation (1) the projection of extracted speech features on to the total variability space [13] can be considered as a probabilistic compression process [14]. That reduces the dimensionality of a channel and speaker dependent super-vector of concatenated Gaussian Mixture Model (GMM) means $M_i$, $i \in I^{h0}$

$$M = m + Tw \quad (1)$$

Where $'m'$ is the speaker and channel independent super vector (which is the gender depended UBM super vector). $T$ is a rectangular matrix of low dimension and $w$ is a random vector having a standard normal distribution.

## 3. MODEL DEVELOPMENT

Let us consider $X$ as a random vector i.e. $X = \{x_1, x_2, x_3, \ldots, x_L\}$ as a set of $L$ vectors, each $x_i$ is a $k$-dimensional feature vectors belong to the one particular acoustic class. $L$ is the number of acoustic classes and the vectors $x_i$ are statistically independent. So the probability of the set $X$ for the $\lambda$ speaker model can be expressed as $\log P(X|\lambda) = \sum_{i=1}^{L} \log P(x_i|\lambda)$ The distribution of vector $x_i$ with the $k$-dimensional components are unknown. It is approximately modeled by a mixture of Gaussian densities, which is a weighted sum of $l \leq k$ component's densities, which can be expressed as $P(x_s|\lambda) = \sum_{i=1}^{l} w_i N(x_s, \mu_i, \Sigma_i)$, here $w_i$ is the mixture weight, where, $1 \leq i \leq l$ and $\sum_{i=1}^{l} w_i = 1$.

Each $N(x_s, \mu_i, \Sigma_i)$ is a $k$ variate Gaussian component density presents as

$$N(x_s, \mu_i, \Sigma_i) = \frac{e^{-\{0.5(x_s - \mu_i)' \Sigma_i^{-1} (x_s - \mu_i)\}}}{(2\Pi)^{k/2} |\sqrt{\Sigma_i}|}$$

$\mu_i$ is the mean vector and $\Sigma_i$ is the covariance matrix. $(x_s - \mu_i)'$ is the transpose of $(x_s - \mu_i)$. In the speaker identification from the set of speakers $\{S_i\}$ where $i$ is countable finite and $X$ is given utterances, if we claim that the utterance produce by the speaker $S_k$ from the set of speakers $\{S_i\}$. So the basic goal is how to valid claim that the speaker $S_k$ makes the utterance $X$. The utterance $X$ is a random variate that follows the Gaussian mixture probability distribution. The claim follows the expression $P(S_k/X)$ presents the probability of the utterances $X$ produced by the speaker $S_k$. So $P(\bar{S}_k/X)$ is the probability that the utterances, $X$ is not produced by the speaker $S_k$. Let, $\bar{S}_k = \bigcup_i S_i - S_k$, is the collection of a large heterogeneous speaker from different linguistics, including both genders and from different zones of the globe. $\bar{S}_k$ can be better approximated as universal model or world model (UBM). It is presented as $\bar{S}_k \approx \omega$ (say). Now the claim will be true according to the rule,

$$\text{If} \quad P(S_k/X) \hbar P(\bar{S}_k/X) \quad (2)$$

then the utterance is produced by $S_k$. Otherwise, the claim is false. So the utterance produced by other speaker, except $S_k$, is a probabilistic prediction about the claim. However, the process cannot predicate certain events, with values 0 or 1. According to the general definition of probability, the process produces highest level of prediction about the claimed speaker with the numeric score values. It is often that this predicted score depends upon the acoustic classes that are obtained from the long step procedure shown in Figure 1. So the extracted feature largely depends on the digitalization of analogue acoustic signal. There are high chances that the probability comparison values of the claimed speaker voice may not be the best or highest value in the interval (0, 1). By the bayes theorem the expression (1) produced

$$\frac{P(X/S_k)P(S_k)}{P(X)} > \frac{P(X/\omega)P(\omega)}{P(X)},$$

Since we assume that $X$ is not silent, then clearly $P(X) \neq 0$. We get,

$$\frac{P(X/S_k)}{P(X/\omega)} > \frac{P(\omega)}{P(S_k)} = \lambda_k. \quad (3)$$

To compact the all possible predictions we consider the log on the both side.

$$\log \frac{P(X/S_k)}{P(X/\omega)} \hbar \log \lambda_k = \lambda,$$ The predicted values indicate how closer the claimed speaker is to the existing speaker's voices after comparison. The predicted values are Gaussian in nature so further compactness can be done on the predicted values by the

$$\text{static} \frac{\frac{P(X/S_k)}{P(X/\omega)} - \mu}{\sigma} > \lambda.$$ $\mu$ is the mean and $\sigma$ is the covariance of the predicted score values of the known speakers voice models.

## 4. COSINE DISTANCE SCORING OF i VECTOR APPROCH

According to Bhattacharyya postulate, Let $p(i)$ and $p'(i)$ represent two multinomial populations for different speaker voice, each consisting of N acoustic classes with respective of the derived probabilities $p(i=1),..., p(i=N)$ and $p'(i=1),..., p'(i)$. Since $p(i)$ and $p'(i)$ represent probability distributions, then $\sum_{i=1}^{N} p(i) = \sum_{i=1}^{N} p'(i) = 1$. So the divergence type measure between the above set of distributions is defined as $\rho(p,p') = \sum_{i=1}^{N} \sqrt{p(i)p'(i)}$. The geometric interpretation as the cosine of the angle between the $N$-dimensional acoustic vector space can be written as $\{\sqrt{p(1)},......,\sqrt{p(N)}\}^T$ and $\{\sqrt{p'(1)},......,\sqrt{p'(N)}\}^T$, where $T$ stand for transpose. If the probability distributions of the acoustic class are identical, we have $\cos\theta = \sum_{i=1}^{N}\sqrt{p(i)p'(i)} == \sum_{i=1}^{N}\sqrt{p(i)p(i)} = \sum_{i=1}^{N} p(i) = 1$

According to the equation (1) we have the expression $M = m + Tw$. Now, $M$ is the conversation side super vector, i-vector is a set of low dimensional total variability factors ($w$) to represent each conversation side. Each factor controls an Eigen dimension of the total variability matrix ($T$). The distance scoring on channel compensated, i-vector $w$ for a pair of conversation sides is the cosine between the target speaker i-vector and the test i-vector. The decision score according to Bhattacharyya postulates presented as,

$$score(w_{target}, w_{test}) = \frac{w_{target} * w_{test}}{\|w_{target}\| * \|w_{test}\|} = \cos(\theta_{w_{target}, w_{test}}) \quad (4)$$

Here, $\theta$ is the decision threshold for accept or reject.

## 5. SIMULATION RESULT AND DISCUSSION

The simulation is done in two stages. In the first stage of simulation we consider the normalized predicted score according to expression (3). The speaker identification is a comparison of the prediction of a said speaker with the number of existing speaker models present in the voice identification system. We make 100 numbers of speaker models with 10 clusters. Each cluster contains 10 speaker's models. We consider three speakers A, B, C for testing purpose. We use initial known voice models of the speakers A, B and C to the three different clusters. The known voice model for the Speaker 'A' is placed into the first cluster with speaker identifier number 1. The known voice model for the Speaker 'B' is placed into the second cluster with speaker identifier number 2. Voice model for a known speaker is placed into the first cluster with speaker identifier number 3 as an impostor speaker of 'C'. The known voice model for the Speaker 'C' is placed into the third cluster with speaker identifier number 4. The other 96 impostor speaker's models are placed along the 10 clusters make the each cluster size 10. At the first stage of identification, we consider two-sample voice of the speaker 'A' and 'B'. These are pure voices of the respective speakers. Figure 3 presents the predicted score values of the simulation. All the figures are presented in the first 6 speaker's identifier matching prediction.

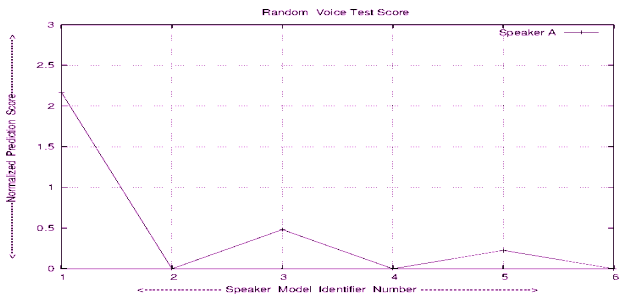

**Figure 3. Random Voice Test Score**

Figure 3 presents the predicted matching value in the list of target models. The new voice sample of the Speaker 'A' match with the model identifier 1 with normalized value 2.2, the predicted normalize score value with the models identifier number 2,3,4,5,6 are 0.0,0.45,0.0,0.34,0.0, respectively. These models belong to different clusters. Figure 4, presents the predicted normalized score value for the new voice sample of the speaker 'B'. The simulated result show the prediction about the new voice sample match with the model identifier number 1,2,3,4,5,6 are (0.0,3.0,0.0,0.0,0.5,0.0) respectively.

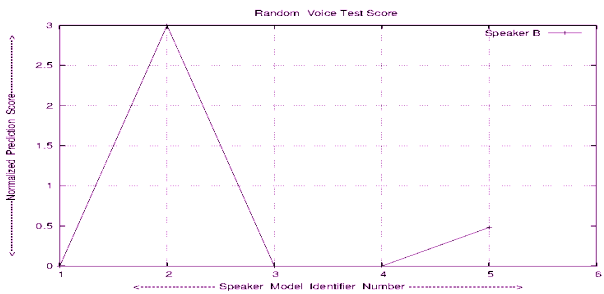

**Figure 4. Random Voice test Score**

The new voice sample matches with the model identifier number 2 with normalized score value 3.0, and also with a model identifier number of 5 with normalized score value 0.5. The above simulation result gives a physical significance of the environment to select the decision threshold of the normalized score values. According to the figure 3 and figure 4, we consider (1.0) as a decision threshold for level of acceptance and rejection to maintain the equal error rate at minimum level. According to this decision threshold, figure 3 shows that the new voice is the voice of the speaker A, and figure 4 shows the new voice is Speaker's B voice.

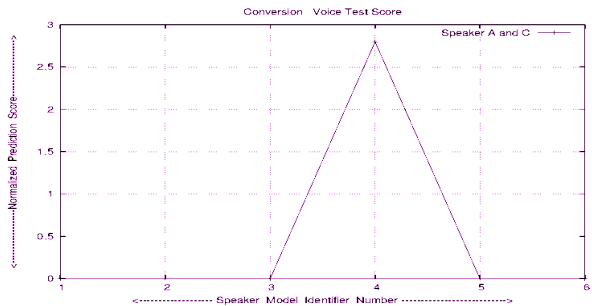

**Figure 5. Conversion Voice Test Score**

Figures 5, 6 and 7 present the predicted score for the conversion voice sample AB, AC and BC. In the target list the speaker voice models 'A', 'B', and 'C' are present. We would like to identify in which conversion speaker 'C' is present. Figure 5 presents normalized predicted score value for the conversion voice sample of the speaker A and C. The predicted normalized score is 2.8 for the model identifier 4. The model identifier 4 is the speaker 'C'. According to the GMM based hypothetical testing prediction about the speaker 'C', is present in this conversion voice sample. The simulation result does not contribute any prediction about the speaker 'A', but truly speaker 'A' is present in the conversion. So it is a false rejection for speaker 'A'. Figure 6 presents the predicted normalized score for the conversion voice sample for speaker 'B' and 'C'. The simulation result shows that the model identifier 2 is present in the conversion with the predicted score value 0.7 and the model identifier 4 is present in the conversion with the normalized score value 2.6. Since we have considered the accepted level of threshold value as (1.0), hence we conclude that model identifier 4 i.e., speaker 'C' is present in this conversion strongly.

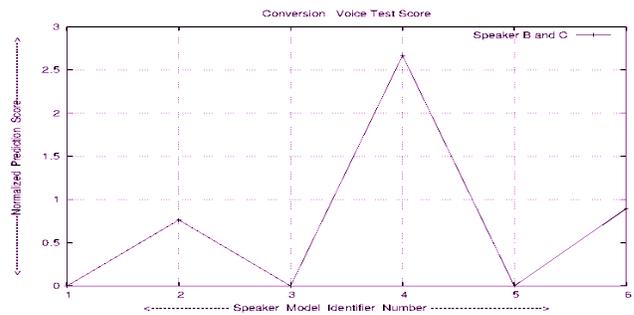

**Figure 6. Conversion Voice Test Score**

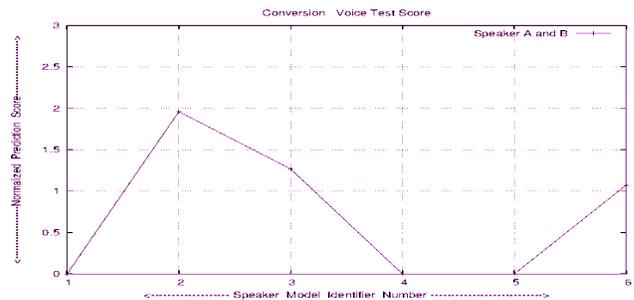

**Figure 7. Conversion Voice Test Score**

The model identifier 2 i.e. speaker 'B' present in the conversion is very less and according to the level of acceptance we ignore the presence, but truly the speaker 'B' is present in the conversion. It is the false rejection. The simulation result for the conversion voice sample of the speakers 'A' and 'B' are presented in the figure 7. The predicted normalized score values shown that the model identifier 2 i.e. speaker 'B' is present in the conversion with the value 2.0. The model identifier 3 i.e. another speaker but not 'A', 'B' and 'C' exist in the conversion with the predicted normalized score 1.3 and the model identifier 6 i.e. a speaker except { A,B,C} match with the score value (1.2). According to the level of acceptance, if we considered 1.5 as a threshold for the environment via figure 3 and figure 4. The simulation result shows that the speaker 'C' is absent in the conversion. Now if we consider the accepted threshold as 1.0, then the prediction that the model identifier number 3 and model identifier number 6 are present in the conversion. These are the impostor models presented in the voice system. Model identifier 3 and model identifier 6 are not the voice model of any of the speakers {A, B, C}. Clearly, it indicates the false accept. The false accept is more vulnerable than false reject. The performance of the identification depends on how much we can reduce the false accept. The second-stage simulation, we used the expression (4). The number of existing models present in the list is 140. There are 10 clusters of speaker models. The target cluster list contains 7 speakers in which 4 speakers are true speakers, and 3 speakers are the impostor's speakers. Speaker ID 1,2,3,4 are true speakers, and speaker ID 5, 6, 7 are imposter speaker models. The model created language and testing language are summarized in table 1.

**Table 1 Languages used in Test and Model**

| Speaker Model ID (voice Collected from distinct native Indian speakers) | Language Used | Test Voice for Speaker ID | Language Used |
|---|---|---|---|
| 1 | English | 1 | Hindi |
| 2 | Bengali | 2 | English |
| 3 | Hindi | 3 | English |
| 4 | Hindi | 4 | English |
| 5 | English | 1 | Oriya |
| 6 | Bengali | | |
| 7 | Hindi | | |

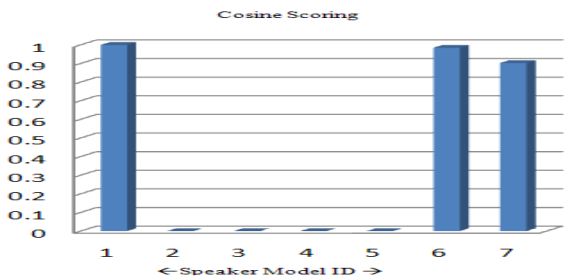

**Figure 8. Cosine Scoring Prediction**

The created model for the speaker ID 1 is in English language, speaker ID 2 created in Bengali language. Figure 8 presents the predicted simulation scores of the input voice of the speaker ID 1 in Hindi language. The predicted cosine score for the test voice is 0.99964 in favour of speaker ID 1 and 0.98453 in favour of speaker ID 6 and 0.9024 in favour of speaker ID 7. Clearly, speaker ID 6 and speaker ID 7 are the false accept. Figure 9 presents the predicted scores 1.000 for true speaker ID 2 and 0.99401 is false accepts for speaker ID 3, 0.9894 is false accepts for the speaker ID 6 also.

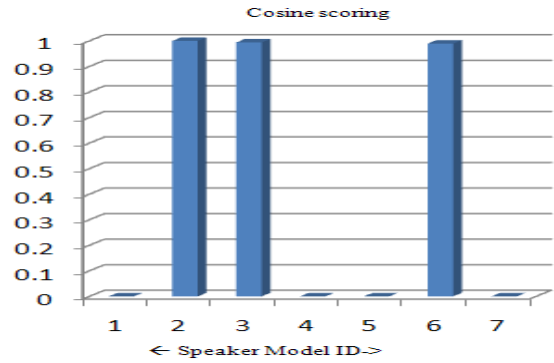

**Figure 9. Cosine Scoring Prediction**

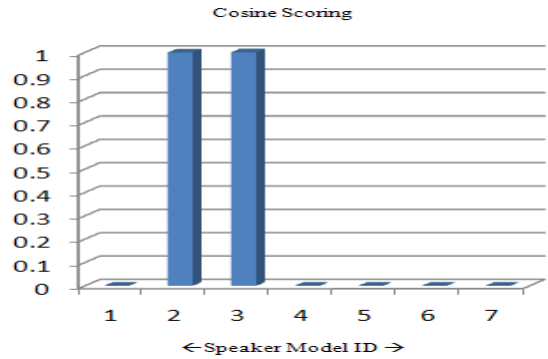

**Figure 10. Cosine Scoring Prediction**

Figure 10 and Figure 11 present the predicted cosine score with the values 0.99999 true accepts for speaker ID 3 and 0.99917 false accepts for the Speaker ID 2. Figure 11 predicted cosine score for the true speaker ID 4 is 1.000 and 0.17997 are false accepts for the speaker ID 3 and 0.46936 is false accepts for the speaker ID 7.

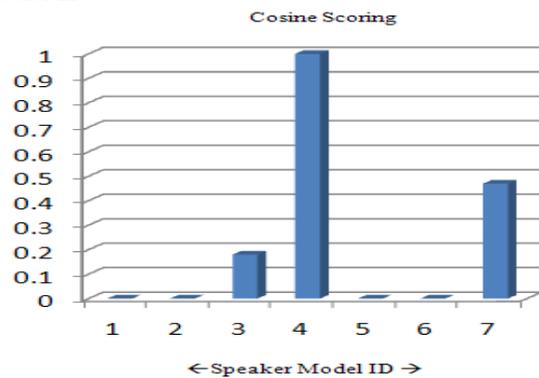

**Figure 11. Cosine Scoring Prediction**

Figure 12 presents the simulated predicted cosine score for the speaker ID 1. The input test voice is in Oriya language. The cosine predictions are 1.000 in favour of speaker ID 1, that is true. 0.87672 Score value for speaker ID 2 is false accept, similarly predicted score 0.9053 and score 0.90343 in favour of speaker ID 6 and 7. Those are false accepting in figure 12.

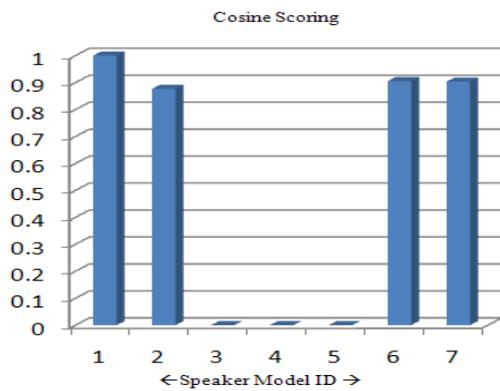

**Figure 12. Cosine Scoring Prediction**

## 6. CONCLUDING REMARKS

This work presents the impact of threshold to identify the vocal tracts. According to figure 3 and figure 4, we choose the decision threshold based on the environment of the collected voice samples. If the environment is very noisy, then obviously the threshold must be reduced. In the case of lesser noisy environments, the decision threshold value must be increased. According to the figure 7 if the decision threshold is considered as 1.0 then obviously the model identifier number 3 i.e. speaker 'C' must be identified as present in the conversion but the speaker 'C' is truly not present in the conversion. It will be a case of false accepts, which will brings the worst impact on the performance of the identification procedure. Conclusion drawn from the simulation result is to first select the decision thrashed from the known speakers according to the environment. The specified threshold applies to identify the unknown speaker through the process. The unknown speaker voice should be collected from the same environment as well as for the GMM model creation voice also. The model voice and test voice should be collected through the same channel to achieve the better system performance. In the second stage of the cosine based predicted scoring, if we consider the highest predicted score for every monolog testing then we get better performance. If we consider the accepted threshold as 0.9, then 2nd best is always a false accepts. In the case of multi speaker voice testing decision control angle $\theta$ is very important for false accepts.

## 7. ACKNOWLEDGMENTS

The author would like to thank Niranjan Kumar Mandal of Vidyasagar University for his continuous encouragement and motivation.